\documentstyle[pre,preprint,aps]{revtex}
\tightenlines
\begin{document}
\draft
\title{Number of distinct sites visited by $N$ random walkers on\\ a
Euclidean lattice}
\author{S. B. Yuste and L. Acedo}
\address{Departamento de F\'{\i}sica, Universidad  de  Extremadura, \\
E-06071 Badajoz, Spain}
\date{\today}
\maketitle
\begin{abstract}
The evaluation of the average number $S_N(t)$ of distinct sites  visited up to time $t$ by $N$ independent random walkers all starting from the same origin on an Euclidean
lattice is addressed. We find that, for the nontrivial time regime and for large $N$,  $S_N(t) \approx \widehat S_N(t) (1-\Delta)$, where $\widehat S_N(t)$ is the volume of a hypersphere of radius $(4Dt \ln N)^{1/2}$, 
$\Delta=\frac{1}{2}\sum_{n=1}^\infty \ln^{-n} N \sum_{m=0}^n
s_m^{(n)} \ln^{m} \ln N$, $d$ is the dimension of the lattice, 
and the coefficients $s_m^{(n)}$ depend on the dimension and time. The
first three terms of these series are calculated explicitly and the resulting expressions are compared with other approximations and with simulation results for dimensions 1, 2, and 3.  Some implications of these results on the geometry of the set of visited sites are discussed.
\end{abstract}

\narrowtext

\section{INTRODUCTION}
\label{sect_1}
Usually, the extremely successful theory of random walks is only concerned with problems that involve a {\em single} ($N=1$) random walker. A solid reason for this is the understanding that the average properties of the single diffusing walker serve to describe the global properties of systems formed by many walkers. However, there are other interesting diffuson problems  that  involve many random walkers for which  the diffusive behavior of {\em every}  walker of the total of $N$ is relevant, i.e., diffusion process that {\em cannot} be described  by averaging over the properties of a single walker \cite{Shlesinger}.  The problem of evaluating the time spent by the first $j$ particles out of a total of $N$ to escape from a given region is a clear example \cite{WSL,PRLEYus}.
Another important example, which is the subject of this paper, is the  problem of evaluating the average number $S_N(t)$ of 
distinct sites visited (or territory explored) by  a set of $N$ independently diffusing random walkers up to time $t$ \cite{Larral1,Larral2}.

The case $N=1$ has been studied in detail  since it was posed by Dvoretzky and Erd\"os \cite{DE} and is discussed in many general references \cite{Montroll,Hughes,Weiss}. 
However, the  multiparticle ($N>1$) version of this  problem has
been systematically treated only after the pioneering works of
Larralde {\em et al.} \cite{Larral1,Larral2}. These authors addressed the problem of evaluating the territory covered by a set of $N$ independent random walkers, all initially placed at the same point, that diffuse with steps of finite variance on Euclidean lattices. They found asymptotic expressions for $S_N(t)$ for $N \gg 1$, and described the existence of three time regimes.
Their results can be summarized as follows:
\begin{equation}
\label{SNtLar}
S_N(t) \sim \left\{ 
\begin{array}{lr}
t^d  & t<t\ll t_{\times} \\
\noalign{\smallskip}
t^{d/2} \ln^{d/2}\left( x \right), & t_{\times} \ll t \ll 
t^{'}_{\times} \\
\noalign{\smallskip}
N S_1(t), &  t_{\times}^{'}\ll t
\end{array} 
\right. \; ,
\end{equation}
where $x=N$ for $d=1$, $x=N/\ln t$ for $d=2$ and $x=N/\sqrt{t}$ for $d=3$
\cite{Larral1,Larral2}. The properties of $S_1(t)$ are well known; in particular, $S_1(t)\sim t^{1/2}$ for $d=1$, $S_1(t)\sim t/\ln t$ for $d=2$ and $S_1(t)\sim t$ for $d=3$.
In the very short-time regime ($t \ll t_{\times}$), or regime I, there are so many
particles  at every site that all the nearest neighbors of the
already visited sites are reached at the next step, so that the number of distinct
sites visited grows as the volume of an hypersphere of radius $t$, $S_N(t) \sim t^d$.
The regime III ($t_{\times}^{'}\ll t$), or long-time regime,  corresponds to the
final stage in which the walkers  move far away from each other
so that their trails (almost) never overlap and $S_N(t) \sim N S_1(t)$.  The crossover time from regime I to regime II is given by $t_{\times} \sim \ln N$ for every lattice.
This can be easily understood if we take into account that the number of
particles on the outer visited sites for very short times will decrease
as $N/z^t$, where $z$ is the coordination number of the lattice, so that the overlapping regime will break approximately when $N/z^t \sim 1$ or,  equivalently, $t_{\times} \sim \ln N$. 
Regime III never appears in the one-dimensional case (i.e., $t_{\times}^{'} \sim \infty$), 
but
$t_{\times}^{'} \sim e^N$ for $d=2$ and $t_{\times}^{'}=N^2$ for $d=3$.
These crossover times will be obtained readily from the
mathematical formalism discussed in the present paper. 
The most interesting regime
is regime II ($t_{\times} \ll t \ll 
t^{'}_{\times}$), or the intermediate regime. 
For this time regime we will obtain explictly the main term and the first  two corrective terms of the asymptotic
expression of $S_N(t)$ for  $N\gg 1$.  Higher corrective terms could be calculated as our method allows  them to be obtained in a systematic way. The contribution of these corrective terms  cannot be ignored even for very large values of $N$ because they decay logarithmically with $N$.
However, as we will see in section \ref{sect_5}, the use of two corrective terms leads to a very good agreement
with simulation results  for relatively small values of $N$ ($N \gtrsim 100$).

The paper is organized as follows. The asymptotic evaluation of $S_N(t)$ for a 
$d$-dimensional Euclidean lattice is discussed in detail in Sec.\ \ref{sect_2}. Some geometric implications of this result are discussed in Sec.\ \ref{sect_3}. In Sec.\ \ref{sect_4}, we compare
our zeroth- (i.e., main), first- and second-order term approximation for $S_N(t)$ with other approximations and with computer simulations for 
one-, two- and three-dimensional simple Euclidean  cubic lattices. The paper ends with some
remarks on the applicability of this method to other diffusion problems and different media.
Some technical details are discussed in an Appendix.

\section{THE NUMBER OF DISTINCT SITES VISITED}
\label{sect_2}
We consider a group of $N$ random walkers starting from an origin site ${\bf r}=0$ at time $t=0$. 
 A survival probability, $\Gamma_N (t,{\bf r})$, is defined
as the probability that site ${\bf r}$ has not been visited by the random
walkers before time $t$. Similarly, we can define a mortality function,
$1-\Gamma_N (t,{\bf r})$, as the probability that site ${\bf r}$ has been visited
by at least one walker in the time interval $(0,t)$. The relationship between the number of distinct sites visited, $S_N(t)$, and the
survival probability is~\cite{Larral1,Larral2}
\begin{equation}
\label{SG}
S_N(t)=\sum_{\bf r}\, \left\{ 1 - \Gamma_N(t,{\bf r}) \right\}\; .
\end{equation}
For independent random walkers, we have $\Gamma_N(t,{\bf r})=\left[
\Gamma_t ({\bf r}) \right]^N$, where $\Gamma_t ({\bf r})\equiv \Gamma_1(t,{\bf r})$ is the one-particle
survival probability. Next, the discrete
analysis implicit in Eq. (\ref{SG}) is  replaced by a continuous one.
Thus, we  write 
\begin{equation}
\label{SGC}
S_N(t)=\int_0^\infty\,\left[ 1 - \Gamma_t^N(r) \right] d\, v_0\, r^{d - 1} d r\; ,
\end{equation}
where $v_0$ is the volume (i.e., the number of lattice sites) of the hyphersphere with unit radius.
It has been found for Euclidean lattices that \cite{Larral2}
\begin{equation}
\label{Gxi}
\Gamma_t(r) \approx \tilde{\Gamma}_t(\xi) = 1 - A \xi^{-2\mu} e^{-d \xi^2/2} \left(
1 + \sum_{n=1}^\infty h_n \xi^{-2 n} \right)\; ,
\end{equation}
for $\xi \equiv r/\sqrt{2 d D t}\gg 1$. Here $D$ is the
diffusion coefficient defined through the Einstein relation
$\left\langle r^2 \right\rangle \approx 2 d D t$, $t \rightarrow \infty$, with  
$\left\langle r^2 \right\rangle$ being the mean-square displacement of a single random walker.
The values of $A$, $\mu$ and $h_1$ for $d=1$, $2$
and $3$ are shown in Table \ref{table1}. A change to the new variable $\xi$
and integration by parts (taking into account that $\tilde{\Gamma}_t(\infty)=1$), yields
\begin{equation}
\label{SGCxi}
S_N(t) = v_0  (2 d D t)^{d/2} J_N (d;0,\infty)\; ,
\end{equation}
where
\begin{equation}
\label{JNab}
J_N(d; a, b)= \int_a^b \, N\,\Gamma_{t}^{N-1}(\xi)
\frac{d \Gamma_t(\xi)}{d \xi} \xi^{d} d \xi\; ,
\end{equation}
In order to evaluate the asymptotic behavior of $J_N(d;0,\infty)$ it is
convenient to make the decomposition
\begin{equation}
J_N(d;0,\infty)=
	J_N(d;0,\xi_{\times})+J_N(d;\xi_{\times},\infty)\; ,
\end{equation}
where $\xi_{\times}$ is a value that should satisfy the following two  conditions:
(a) $\xi_{\times}$ is large enough so that $\Gamma_t(r)$ can be
well approximated by its asymptotic approximation $\tilde\Gamma_t(\xi)$ for $\xi \ge \xi_{\times}$, and (b) small enough
so that 
\begin{equation}
\label{defxitimes}
\tilde\Gamma^N_t(\xi_{\times})=1/N^p ,
\end{equation}
with $p > 1$ (say $p=2$). From Eq. (\ref{Gxi}) it is straightforward to see that
\begin{equation}
\label{xitimes}
\xi_{\times}^2 \sim \ln N
\end{equation}
satisfies both conditions.
On the other hand, because at most $d\Gamma/d\xi={\cal O}(1)$, and $\Gamma_t(\xi)$ is a monotonous growing function, $J_N(d;0,\xi_\times)$ is bounded by a term that goes as $N \tilde \Gamma_t^{N-1}(\xi_\times) \xi_\times^{d}$, or equivalently, from  Eq.\ (\ref{defxitimes}), by a term that goes mainly as $N^{1-p}$. But shortly we will show that $J_N(d;\xi_{\times},\infty)$
goes essentially as $\ln^{d/2} N$;
this means that
$J_N(d;0,\xi_{\times})$ is asymptotically smaller than any term in the
asymptotic expansion for $J_N(d;0,\infty)$ and thus we can write
\begin{equation}
\label{JNdec}
J_N(d;0,\infty) \approx J_N(d;\xi_{\times},\infty)\; , \quad  N \gg 1\; .
\end{equation}

The previous discussion is illustrated in Fig.\ \ref{figxitimes} for the one-dimensional case. In this figure we have represented
the integrand of $J_N(1;0,\infty)$  
for increasing values of $N$ using as survival probability the exact value $\Gamma_t(\xi)=\text{erf}(\xi/\protect{\sqrt{2}})$   \cite{Hughes,Weiss} and the asymptotic expression of Eq.\ (\ref{Gxi}) up to first order ($n=1$).
Notice that the area below the solid [broken] curve is just the exact [asymptotic approximate]  value of $S_N(t)/(8 D t)^{1/2}=J_N(1;0,\infty)$. 
The value of $\xi_\times$ as given by Eq.\ (\ref{defxitimes}) with $p=2$ is marked with a symbol. It is clear from the figure that, for large $N$, (a) the integrand of $J_N(1;\xi_\times\infty)$ is well represented by the asymptotic expression of $\Gamma_t(\xi)$, and (b) that, as stated for the general case in Eq.\ (\ref{JNdec}), $J_N(1;0,\xi_\times) \ll J_N(1;\xi_\times\infty) \approx J_N(1;0,\infty)$.

 From Eq. (\ref{Gxi}), one easily finds that
\begin{equation}
\label{Gdev}
\frac{d\tilde{\Gamma}_t(\xi)}{d \xi}
\left[ 1 - \tilde{\Gamma}_t(\xi)\right]^{-1}=
2 \xi \sum_{n=0}^\infty \, j_n \xi^{-2 n}\; ,
\end{equation}
with $j_0=d/2$, $j_1=\mu$, $j_2=h_1$, \ldots  By inserting Eq.\ (\ref{Gdev}) into Eq.\ (\ref{JNab}) one has the following expansion for $J_N(d;\xi_{\times},\infty)$:
\begin{equation}
\label{JNexp}
J_N(d;\xi_{\times},\infty) \approx 2 N
\sum_{n=0}^\infty j_n K_{N-1} \left(d-2n+1 \right)\; ,
\end{equation}
with
\begin{equation}
\label{KN}
K_N(\alpha)=\int_{\xi_{\times}}^\infty \, \xi^\alpha \tilde{\Gamma}_t^N(\xi)
\left[ 1-\tilde{\Gamma}_t(\xi) \right] d \xi \; .
\end{equation}
By means of the substitution
\begin{equation}
\label{zdef}
\tilde{\Gamma}_t(\xi)=e^{-z}\; ,
\end{equation}
we get a more convenient expression for $K_N(\alpha)$: 
\begin{equation}
\label{KNz}
K_N(\alpha)=\int_0^{z_{\times}} \, e^{-N z} \left( e^{-z}-1 \right) 
\xi^\alpha
\frac{d \xi}{d z} d z\; ,
\end{equation}
where, from Eq.\ (\ref{defxitimes}), $z_x \sim \ln N/N$. The integral in (\ref{KNz}) is of Laplace type but
it is not possible to use Watson's lemma directly to get its asymptotic
behavior because $\xi^\alpha (d\xi/d z)$ has a logarithmic singularity
at $z=0$ \cite{Wong}. The evaluation of $K_N(\alpha)$ requires the inversion of (\ref{zdef})
to obtain $\xi(z)$. By using Eqs.\ (\ref{Gxi}) and (\ref{zdef}) we get
\begin{equation}
\label{zdefi}
-\frac{d}{2} \xi^2+\ln A+\mu \ln \xi^{-2}+\ln \left( 1+\sum_{n=1}^\infty \, h_n
\xi^{-2 n} \right)=\ln \left( 1-e^{-z} \right)\; .
\end{equation}
The function $\xi(z)$  can be readily obtained from  this equation to first approximation: Notice that, as long as 
\begin{equation}
\label{xicondi}
\xi^2\gg |\ln A|,
\end{equation}
the left hand side of Eq.\ (\ref{zdefi}) can be approximated by $-d \xi^2/2$, so that the first-order solution to 
Eq.\ (\ref{zdefi}) is $\xi^2(z)\approx -2\ln[1-\exp(z)]/d$. 
Equation (\ref{zdefi}) can be systematically solved in order to get higher-order approximations (see Appendix). The result is 
\begin{equation}
\label{xiz}
\xi=x^{-1/2} \sum_{n=1}^\infty \delta_n x^{n}\; ,
\end{equation}
where $x=-(d/2)/\ln[1-\mbox{exp}(-z)]$. 
The substitution of  Eq.\ (\ref{xiz}) into Eq.\ (\ref{KNz}) (see Eq.\ (\ref{xiadxi}) in 
the Appendix) yields
\begin{equation}
\label{KNzexp}
K_N(\alpha)=\sum_{n=0}^\infty\sum_{m=0}^n 
\frac{2^{\left(\alpha-1\right)/2}}{ d^{\left(\alpha+1\right)/2}} k_m^{(n)} I\left(\frac{\alpha}{2}-n-\frac{1}{2},m;N\right),
\end{equation}
where
\begin{equation}
\label{Idef}
I(n,m;N)\equiv \int_0^{z_{\times}} dz e^{-N z} (-\ln z)^n \ln^m(-\ln z).
\end{equation}
The evaluation of $I_N(n,m;N)$ for $N\rightarrow \infty$ has been discussed in \cite{WSL,Wong}. For the sake of completeness, we give here explicitly their expressions up to the order required to find  $S_N(t)$ to second order in $1/\ln N$:
\begin{eqnarray}
I(n,0;N)&\approx &\frac{1}{N}\ln^n N
\left[1+\frac{n \gamma}{\ln N}+\frac{n(n-1)}{2} \frac{\gamma^2+\pi^2/6}{\ln^2N} +\cdots \right] ,\\
I(n,1;N)&\approx &\frac{1}{N}\ln^n N
\left[\ln \ln N \left(1+\frac{n\gamma}{\ln N}\right)+ \frac{\gamma}{\ln N}
 +\cdots \right] ,\\
\label{In2N}
I(n,2;N)&\approx & \frac{1}{N}\left(\ln^n N\right) \ln^2\ln N +\cdots
\end{eqnarray}
where $\gamma \simeq 0.577215$ is the Euler constant. 
 Using these results we get from Eqs.\ (\ref{SGCxi}), (\ref{JNexp}) and (\ref{KNzexp}) the following expansion for the average number of distinct sites visited on a Euclidean lattice of dimension $d$:
\begin{equation}
\label{SNt}
S_N(t) \approx \widehat S_N(t) (1-\Delta)
\end{equation}
with
\begin{eqnarray}
\label{SNtgorro}
\widehat S_N(t)&=&v_0 \left( 4 D t \ln N \right)^{d/2},  \\
\label{Delta}
\Delta  \equiv  \Delta(N,t) & = &
\frac{1}{2}\sum_{n=1}^{\infty} \ln^{-n} N \sum_{m=0}^n s_m^{(n)} \ln^{m} \ln N  \; 
\end{eqnarray}
and where, up to second order ($n=2$),
\begin{eqnarray}
\label{s10}
s_0^{(1)}&=& -d \omega  ,\\
 s_1^{(1)}&=& d \mu   ,\\
 s_0^{(2)}&=&
d\left(1-\frac{d}{2}\right) \left( \frac{\pi^2}{12}+\frac{\omega^2}{2} \right) 
-d\left(\frac{d h_1}{2}-\mu \omega\right) ,\\
 s_1^{(2)}&=&  - d \left(1-\frac{d}{2}\right)\mu \omega - d\mu^2 ,\\
s_2^{(2)}&=& \frac{d}{2} \left(1-\frac{d}{2}\right) \mu^2  .
\label{scoef}
\end{eqnarray}
Here $\omega=\gamma+\ln A+ \mu \ln(d/2)$, and  $A$, $\mu$ and $h_1$ are given in Table \ref{table1} for $d=1$, $2$ and $3$.
Notice that the time dependence of $\Delta(N,t)$ comes from the term $\omega$ through the function $A(t)$. However,  this function does not depend on time for the one-dimensional case and thus $\Delta$ only depends on $N$.

Recently, Sastry and Agmon \cite{SA} have found an approximate formula for $S_N(t)$ for the one-dimensional case. 
The  straightforward method used by these authors is based on the fact that the function $\Gamma_t^N(r)$ that appears in the integrand of Eq.\ (\ref{SG}) approaches a step function when $N\rightarrow \infty$. In this way they found 
\begin{eqnarray}
 S_{N}(t) &\approx &
4\sqrt{Dt}\sqrt{\ln N-\ln\sqrt{\alpha \ln N}} 
\label{SntSastry1}\\
&\approx& 4\sqrt{Dt\ln N}
    \left[1-\frac{1}{4}\frac{\ln\ln N-\ln \alpha}{\ln N}+
		{\cal O}\left(\frac{\ln^2\ln N}{\ln^2 N}\right)\right]
\label{SntSastry2}
\end{eqnarray}
where $\alpha$ is given by $\alpha=\pi\exp(-2/\pi)\simeq 1.66$.  
It is instructive to compare this formula with the first-order approximation of Eq.\ (\ref{SNt}) for the one-dimensional case:
\begin{equation}
S_{N}(t) \approx 
4 \sqrt{Dt\ln N}
    \left[1-\frac{1}{4}\frac{\ln\ln N-2\omega}{\ln N}+
		{\cal O}\left(\frac{\ln^2\ln N}{\ln^2 N}\right)\right],
\label{SNt1d}
\end{equation}
Note that the prefactor $4(Dt\ln N)^{1/2}$ of the formula of Sastry and Agmon is in agreement with that of Eq.\ (\ref{SNt1d}). In Ref.\ \cite{SA},  they found it  ``amusing'' that the value $\alpha=1$ produces  very good agreement between the approximation of Eq.\ (\ref{SntSastry1}) and the exact numerical integration. Our Eq.\ (\ref{SNt1d}) enlightens this point: Comparing  Eqs.\ (\ref{SntSastry2}) and  Eq.\ (\ref{SNt1d}) one sees that $\ln \alpha$ is playing the role of $2\omega$. But $\omega=\gamma-\frac{1}{2}\ln \pi=0.0048507\cdots$ for the one-dimensional lattice, so that $\ln \alpha$ when $\alpha=1$ leads to a good approximation to the rigorous coefficient $2 \omega$. The equation of Sastry and Agmon for $\alpha=1$ and our first-order approximation should thus be very close. This is clearly confirmed in Fig.\ \ref{figSNt}.

A question to be answered is why  Eq.\ (\ref{SNt}) is valid for time
regime II only, i.e., why it is not always valid for arbitrarily large
values
of time.  The reason is that our formulas have been obtained by assuming
that the condition (\ref{xicondi}) holds for those values of $\xi$ which are inside the integration interval $\left[ \xi_{\times},\infty \right]$ of the relevant integral  $J_N(d;\xi_{\times},\infty)$ that is responsible for  the asymptotic behavior of $S_N(t)$.  This implies that for our procedure to work, it is necessary that $\xi_{\times}^2 \gg \ln A$ or, from  Eq.\ (\ref{xitimes}), that 
\begin{equation}
\label{NvsAcon}
\ln N \gg |\ln A|  .
\end{equation}
Thus we can estimate the time $\tau_\times$ for which our
method breaks down by solving $|\ln A(\tau_\times)|\sim \ln N$. From the
expressions for $A$ quoted in Table \ref{table1} one finds that
$\tau_\times \sim e^N$ for $d=2$ and $\tau_\times\sim N^2$ for $d=3$. For $d=1$ and large $N$, the condition (\ref{NvsAcon}) always holds because $A=(2/\pi)^{1/2}$ is a constant and then $\tau_\times = \infty$.
We see that the upper times $\tau_\times$ beyond which Eq.\ (\ref{SNt}) is no longer valid coincide with the crossover times $t_{\times}^{'}$ defined in Sec. \ref{sect_1}, so we can
say that the expressions for $S_N(t)$ given in this paper are valid only in the time regime II. This means that our procedure marks its own limit of validity as that of regime II, thus predicting the existence of a crossover time in a natural way, i.e., as a consequence of the mathematical formalism.

\section{GEOMETRIC PROPERTIES OF THE EXPLORED REGION}
\label{sect_3}
In this section we will give a geometric interpretation of the main result
of this paper, namely, Eq.\ (\ref{SNt}). The quantity $S_N(t)$ is by definition the volume of the
region $\Omega$ explored by $N$ random walkers after a time $t$ from their  initial
deposition on a given site of the lattice (if the length of the lattice
bonds is taken as the unit). For very short times (regime I or
$t \ll \ln N$) the exploration is performed in a compact way because all the
neighbor sites of any visited site are always visited at the next
time step. Therefore, the explored region $\Omega$ is an hypersphere whose radius grows
ballistically and its volume is proportional to $t^d$.
After the regime II is reached, the development of two qualitatively different
zones in the explored volume is observed: (i) an hyperspherical compact core of
visited sites, and (ii) a corona of dendritic nature characterized by filaments created by those relatively few walkers that are wandering in the outer regions, i.e., wandering at distances significantly larger than the root-mean-square displacement $\langle r^2 \rangle^{1/2}=\sqrt{2dDt}$ of a single walker.
Figure \ref{figSnapshot} shows a snapshot of the set of sites  visited by $N=1000$ random walkers at time $t=900$ (every walker makes a jump at each time unit) for dimension two. The visited sites are in white and the inner black and outer white circles delimit the corona.
The radius $R_{+}$ of the outer circle is equal to the maximum displacement from the origin reached by any of the walkers at time $t$. 
It has been argued in \cite{RC} that the volume of this outer circle is on average given by the main term of (\ref{SNt}), i.e., by $\widehat S_N(t)=v_0(4Dt\ln N)^{d/2}$. From this statement we can draw two conclusions: First, that the average radius of this outer circle is 
\begin{equation}
\label{R+}
R_{+}\approx \left(4Dt\ln N \right)^{1/2} ,
\end{equation} 
and second, that the asymptotic corrective terms (given by $\Delta$) to $S_N(t)$ account for the number of {\em unvisited} sites that are inside the corona. In other words, $\Delta$ is the fraction of the volume inside the external circumference that has not been visited by any of the $N$ random walkers. This result can be used \cite{RC} to easily estimate that the thickness of the dendritic corona is approximately 
given by $R_{+} \Delta$.

It is also noteworthy that $\Delta(N,t)$ depends on $t$ very smoothly in the time regime II as this dependence is due to terms proportional to powers of $\ln A(t)$ [and $A(t)$ does not change exponentially: see Table \ref{table1}].  For the two-dimensional case, this statement is especially valid because $A(t)\sim 1/\ln t$.  Therefore, the ratio (given by $\Delta$) between the radial size of the corona of $\Omega$ and the radial size of $\Omega$ itself remains almost constant throughout the time regime II. This implies that a conveniently scaled sequence of snapshots of the
set of visited sites should be very similar (in a statistical sense), i.e., we find that $\Omega$ grows, to a large extent, in a self-similar way inside time regime II.  This property is illustrated in Fig.\ \ref{fig:selfsimi}. As Eq.\ (\ref{R+}) shows, the appropriate scale factor must be proportional to $\sqrt{t}$. 
This ``almost'' self-similar behavior disappears as the regime III is approached because the correction to the main term of $S_N(t)$ becomes
as large as this main term, i.e., because $\Delta(N,t)$ approaches the value 1. This transition takes place when $t \approx \tau_\times$ as follows from (\ref{Delta}), i.e., this value coincides with the threshold for regime III deduced in the previous section. From the geometric point of view this transition corresponds to the breaking of the self-similar growing behavior by the appearance of a corona of filaments as large as the compact core, which
finally gives rise to a set of separated trails that (almost) never more overlap. 
For the two-dimensional case the transition time from regime II to regime III is so great for any significant number of walkers that it can not be studied by numerical simulation.

\section{NUMERICAL RESULTS}
\label{sect_4}
We carried out numerical simulations for the number of distinct
sites visited by $N=2^m$, with $m=0,1,\ldots,14$ in two and three
dimensions. For the one-dimensional case it is not necessary to carry out simulations because the survival probability is exactly known on this lattice, 
$
\Gamma_t({\bf r})=\text{erf}\left(\xi/\sqrt{2}\right) ,
$
and therefore the integral for $S_N(t)$ as given by Eq. (\ref{SGC}) can be computed numerically. 

In our simulations, the random walkers are placed initially at the center of an hypercubic box of side $L$.
The regime II is reached almost immediately with
the number of random walkers we have used ($t_{\times} \approx 10$ for
$N=2^{14}$). The simulations were carried out only to a maximum time
$t=200$ which is sufficient for the stabilization of regime II conditions.
The square box side for $d=2$ was taken to be $L=400$ to avoid any
random walker reaching the edge of the box before the maximum time
$t=200$. Memory limitations forced us to reduce the box side to $L=200$ for the three-dimensional case. While this implies a possible appearance
of finite-size effects, we can consider them as to be negligible because the average displacement of the random walkers at the maximum time is small
compared with $L/2$. Each experiment was repeated $10^4$ times in order to achieve reasonable statistics.

\begin{figure}
\caption{
${\bf S}=[S_N(t)/v_0]^{2/d}/(4 d D t \ln N)$ versus $1/\ln N$ for, from top to bottom, dimension $1$, $2$ and $3$ and $t=200$ (inside time regime II). We have used $N=2^m$ with $m=3,\cdots,14$ for $d=2,3$, and   $m=3,\cdots,30$ for $d=1$.
The numerical results are plotted
as filled circles and the broken [solid] lines correspond to the theoretical 
predictions for $S_N(t)$  to first [second] order as given by Eq.\ (\protect{\ref{SNt}}).
Notice that the  approximation of order 0 would be a horizontal line (not shown here) passing through $1/d$.
The crosses correspond to the Sastry and Agmon result of Eq.\  (\protect{\ref{SntSastry1}}) with $\alpha=1$.
The dotted lines correspond to the result of Larralde {\em et al.} given by Eq.\ (\protect{\ref{SNtLar}}) in which the corrected amplitude of the main term has been used (see \protect{\cite{RC}}).
}
\label{figSNt}
\end{figure}
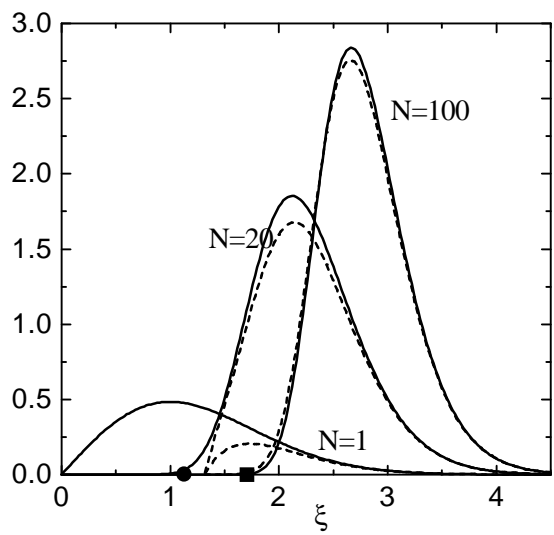
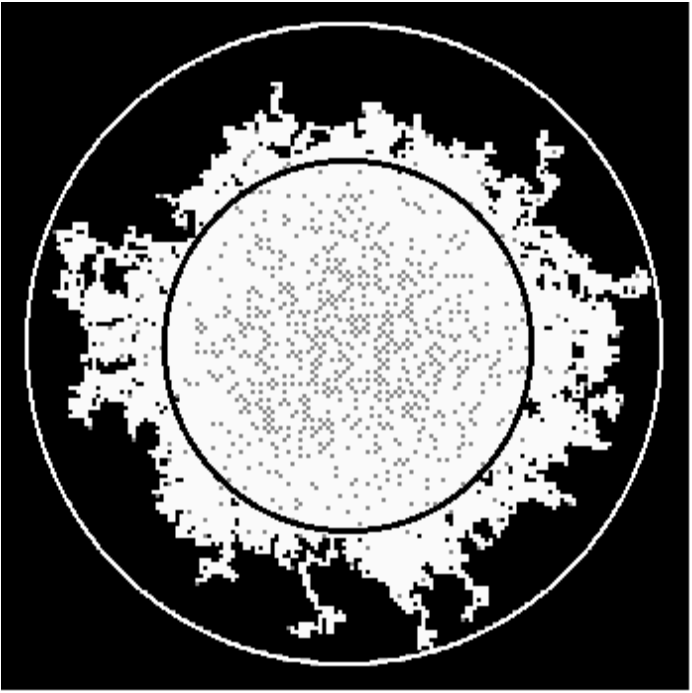
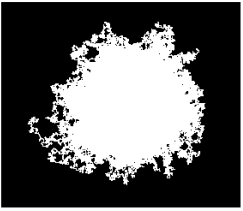
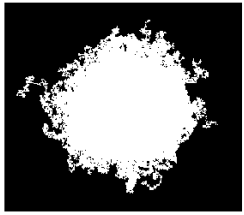
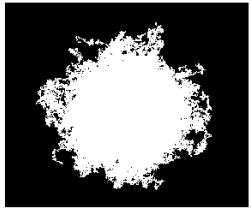
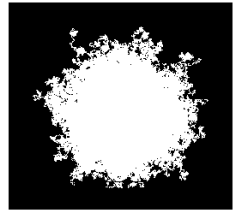
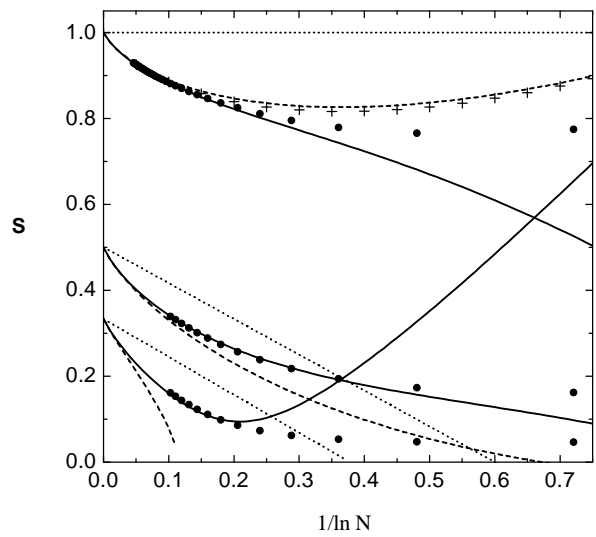

Results are plotted in Fig.\ \ref{figSNt} for one, two and three dimensions.
The dots are the simulation results (numerical integration results in one
dimension) and the broken and solid lines are the prediction
of Eq. (\ref{SNt}) to first and second order, respectively. 
The crosses are the results of Sastry and
Agmon \cite{SA} given by Eq.\ (\ref{SntSastry1}) with $\alpha=1$.
 The dotted lines correspond to the result of Larralde {\em et al.} given by Eq.\ (\ref{SNtLar}) using the correct amplitude of the main term (see \cite{RC}).  
The quantity plotted  is
\begin{equation}
{\bf S}\equiv\frac{1}{d}\left[\frac{S_N}{\widehat S_N}\right]^{2/d} \; ,
\end{equation}
versus $1/\ln N$. From Eq.\ (\ref{SNt}) one sees that the theoretical prediction for this quantity is $ {\bf S}\approx (1/d)(1-\Delta)^{2/d}$. The agreement between the second-order approximation and
the simulations is found to be excellent for $N \gtrsim 100$. Good agreement for lower values of $N$ would be  expected if higher-order terms in the series were included. 
The importance of the corrective terms is evident. For example, for the one-dimensional case, we would need to use values of $N$ as large as $10^{25}$ in order to obtain the same precision with the main term as we get with the main and two corrective terms for values of $N$ as small as $2^6$. Similar statements can be made for the other lattices, as Fig.\ \ref{figSNt} shows.

\section{REMARKS}
\label{sect_5}
In this paper we have developed a method for calculating the mean
number of distinct sites visited by $N$ {\em independent} random walkers on
Euclidean lattices.  The method allows the systematic calculation of the main and corrective asymptotic terms to any order for large $N$. These corrective terms are generally non-negligible as they (essentially) decay as powers of $1/\ln N$. However, we found that the main and first two  corrective terms lead to reasonably good results when relatively small values of $N$ are used (say, for $N \gtrsim 2^7$).  In Sec.\ \ref{sect_3} we proposed a geometric meaning for the main and corrective terms: the main term would account for the volume of the set of visited sites if the exploration of the random walkers were compact, and the corrective terms just improve this  rough estimate because, in the outer regions, the exploration performed by the (relatively few) random walkers that move there is really not compact, thus leading to the formation of a non-compact (a dendritic) external  ring in the set of visited sites.  We hope the above results and ideas could serve as a basis  to gain insight into  problems with {\em interacting} random walkers.

The method developed here for calculating $S_N(t)$ is also useful for evaluating other statistical quantities related to the diffusion of a set of independent random walkers. An example is the number $S_{N+}(t)$ of sites visited by $N$ random walkers on an one-dimensional lattice along a {\em given} direction \cite{SA}. It turns out that the moments (of arbitrary order) of $S_{N+}(t)$ can be readily obtained through a slight modification of Eq.\ (\ref{SNt}).  Another example is the first passage time $t_{1,N}(r)$ to a distance $r$ of the first random walker of a set of $N$. 
First passage times are
relevant statistical quantities in the study of 
diffusion processes  where the arrival of the first particles at a given site produces a significant effect (a ``trigger'' effect). 
These quantities have been calculated for one dimension \cite{PRLEYus,wsl83} (and for some classes of fractals \cite{PRLEYus}) but little is known for dimensions greater than one \cite{WSL}. 
The approximate compact
form of the set of visited sites allows one to estimate the first passage
time via the relation $S_N(t_{1,N}(r)) \approx v_0 r^{d}$ \cite{RC}, which
 means geometrically that we consider the region inside the hypersphere of
radius $r$ where a random walker has arrived by time $t_{1,N}(r)$ as
completely visited (a compact exploration in the sense of de Gennes \cite{Gennes}). Results on  $S_{N+}(t)$ and $t_{1,N}(r)$ obtained using the above ideas will be reported elsewhere.  

The function $S_N(t)$ we have studied is indeed an important quantity concerning the diffusion of $N$ independent random walkers but
there are still many open questions in this problem. One can think, for example, of the absorption probability of the set of $N$ random walkers on a lattice with a random distribution of point-like traps. 
This problem can be formulated in terms of the moments of the number of distinct sites visited by the set of $N$ walkers.
A prediction for the variance of the number of visited sites is a
necessary requisite to tackle this interesting problem as the first-order
approximation based only on its first moment, i.e., on $S_N(t)$, seems to be very imprecise \cite{Hughes}.  As no relationship is known for moments of order higher than one, the absorption problem remains unsolved. 

Finally, it should be pointed out that the expression for $S_N(t)$ given in this paper can be extended to fractal media with some slight changes. We are currently running simulations 
for deterministic (Sierpinski gasket) and stochastic (percolation
aggregate) fractals. Results for these substrates will be published 
elsewhere.

\acknowledgments
Partial support from the DGICYT (Spain) through grant no PB97-1501 and
from the Junta de Extremadura-Fondo Social Europeo through grant IPR99C031 
is acknowledged.

\appendix
\section*{}
\label{appenA}
 We will show in this Appendix how to get Eq.\ (\ref{KNzexp}) from Eq.\ (\ref{KNz}). 
Let us start by showing that the solution $\xi(z)$ of
Eq.\ (\ref{zdefi}) for $z \rightarrow 0$ has the form given
in Eq.\ (\ref{xiz}). For simplicity of notation we will write $u=\xi^{-2}$, $\phi=1-\mbox{exp}(-z)$ and $c=d/2$.
Hence, Eq. (\ref{zdefi}) takes the form
\begin{equation}
\label{phidefi}
-\frac{c}{u}+\mu \ln u+\ln A+\ln\left(1+\sum_{n=1}^\infty\, h_n u^n\right)=\ln \phi\; .
\end{equation}
In the limit $z \rightarrow 0$, it is clear that $u \rightarrow 0$ and $\phi \rightarrow 0$.
This means that, as long as $1/u\gg |\ln (A)|$, the first term on the right-hand side of (\ref{phidefi}) is the
most divergent one so that, as a first approximation, we have
\begin{equation}
\label{fapp}
u\approx -\frac{c}{\ln \phi} \equiv x \; .
\end{equation}
This first-order approximation was already obtained in Sec.\ \ref{sect_2} [see below Eq.\ (\ref{xicondi})]. A better approximation is achieved by writing $u=x(1+\epsilon)$, with $\epsilon$ an small quantity. The substitution of this expression  in (\ref{phidefi}) yields
\begin{equation}
\label{phiep}
\epsilon-\epsilon^2+\frac{\mu x}{c}\ln x+\frac{x}{c} \ln A+\frac{\mu x}{c}\epsilon
-\frac{\mu x}{2 c} \epsilon^2+\frac{h_1 x^2}{c}+\frac{h_1 x^2 \epsilon}{c}+\ldots=0\; ,
\end{equation}
where (\ref{fapp}) has been taken into account. This equation can be  solved by writing $\epsilon$ as 
\begin{equation}
\epsilon=\sum_{n=1}^\infty \epsilon_n x^n \; ,
\end{equation}
and inserting it in (\ref{phiep}). We thus find the following
values for $\epsilon_n$ up to $n=2$:
\begin{eqnarray}
\label{epsilon}
\epsilon_1&=&-\frac{1}{c}\ln\left( A x^\mu \right) ,\\
\noalign{\smallskip}
\epsilon_2&=&\frac{1}{c^2}\ln^2\left( A x^\mu \right)+\frac{\mu}{c^2} \ln
\left( A x^\mu \right)-\frac{h_1}{c} \; .
\end{eqnarray}
Therefore
\begin{equation}
\label{xizapp}
\xi(z)=u^{-1/2}=x^{-1/2}  \left(1+\epsilon\right)^{-1/2}=x^{-1/2} 
\sum_{n=0}^\infty \delta_n x^n  \; ,
\end{equation}
where $\delta_0=1$ and 
\begin{eqnarray}
\label{deltan}
\delta_1 &=& \frac{\ln \left( A x^\mu \right)}{2 c} ,\\
\noalign{\smallskip}
\delta_2 &=&- \frac{1}{8 c^2} \left[\ln^2 \left( A x^\mu \right)+4 \mu
\ln \left( A x^\mu \right)-4 c h_1 \right] \nonumber \; .
\end{eqnarray}
The evaluation
of the integral for $K_N(\alpha)$ in (\ref{KNz}) requires the expression
of $\xi^\alpha d\xi/d z$ as a function of $z$. From Eq.\ (\ref{xizapp}) and taking into account that
$d \xi/dz= (d\xi/dx)(dx/dz)$ and $dx/dz=[x^2/(c\phi)]d\phi/dz$, 
we find that 
\begin{equation}
\label{xiadxi}
\xi^\alpha \frac{d \xi}{d z}=-\frac{1}{2c \phi} \frac{d \phi}{d z} x^{(
1-\alpha)/2}
\left[
	1+\sum_{n=1}^\infty x^n
	 \sum_{m=1}^n \hat{k}_m^{(n)} \ln^{m} \left( A x^\mu \right)
\right] \; ,
\end{equation}
where the coefficients $\hat{k}_m^{(n)}$, $m=0,\ldots,n$
for $n=1$, $2$ are
\begin{eqnarray}
\hat{k}_0^{(1)}&=&-\frac{\mu}{c} \nonumber ,\\
\noalign{\smallskip}
\hat{k}_1^{(1)}&=&\frac{\alpha-1}{2 c} \nonumber ,\\
\noalign{\smallskip}
\hat{k}_0^{(2)}&=&\frac{(\alpha-3) h_1}{2 c}+\frac{\mu^2}{c^2} ,\\
\noalign{\smallskip}
\hat{k}_1^{(2)}&=&\frac{\mu (2 - \alpha)}{2 c^2} \nonumber ,\\
\noalign{\smallskip}
\hat{k}_2^{(2)}&=&\frac{\alpha (\alpha - 4) + 3}{8 c^2} \nonumber \; .
\end{eqnarray}
Let us use $\widehat K_N(\alpha,z)$ to denote the integrand of Eq.\ (\ref{KNz}), i.e.,
\begin{equation}
\label{KNalpha}
 K_N(\alpha)=\int_0^{z_\times} \widehat K_N(\alpha,z)  dz.
\end{equation}
Then, from Eq.\ (\ref{xiadxi}),
\begin{equation}
\label{KNserie}
 \widehat K_N(\alpha,z) =\frac{1}{2c}e^{-Nz} e^{-z} x^{(1-\alpha)/2} 
\left[
	1+\sum_{n=1}^\infty x^n
	 \sum_{m=1}^n \hat{k}_m^{(n)} \ln^{m} \left( A x^\mu \right)
\right] \; .
\end{equation}
Writing  
$e^{-z}=1+{\cal O} (z)$,
 $x=-(c/\ln z) [1+{\cal O} (z/\ln z)]$ and 
$\ln(A x^\mu)=\ln A -\mu \ln (-\ln z)+\mu \ln c+{\cal O} (z/\ln z)$, 
Eq.\ (\ref{KNserie}) becomes
\begin{equation}
\label{KNserieFin}
 \widehat K_N(\alpha,z) =[1+{\cal O} (z)] \frac{1}{2c^{(\alpha+1)/2}}
e^{-N z} (-\ln z)^{(\alpha-1)/2} 
\sum_{n=0}^\infty  \sum_{m=0}^n
k_m^{(n)} (-\ln z)^n \ln^m(\ln z)
\end{equation}
where the coefficients $k_m^{(n)}$ up to second order ($n=2$) are
\begin{eqnarray}
k_0^{(1)}&=&(\alpha-1)  \frac{\omega}{2}-\mu \nonumber , \\
\noalign{\smallskip}
k_1^{(1)}&=&(1-\alpha) \frac{\mu}{2} \nonumber ,\\
\noalign{\smallskip}
k_0^{(2)}&=& (3-\alpha)(1-\alpha) \frac{\omega^2}{8}
+\mu (2-\alpha) \omega+\mu^2+\frac{h_1 c}{2}\left(
\alpha-3\right) \nonumber ,\\
\noalign{\smallskip}
k_1^{(2)}&=&
	\mu\left[ (\alpha-2) \mu+ (\alpha-3)(1-\alpha)\frac{\omega}{4} \right] \nonumber ,\\
\noalign{\smallskip}
k_2^{(2)}&=&\frac{\mu^2}{8} (\alpha-3)(\alpha-1) \nonumber \; ,
\end{eqnarray}
and  $\omega=\gamma+ \ln A+\mu \ln c$. 
Finally, inserting Eq.\ (\ref{KNserieFin}) into Eq.\ (\ref{KNalpha}) we get Eq.\ (\ref{KNzexp}). It should be noted that we have approximated the factor $1+ {\cal O}(z)$ of Eq.\ (\ref{KNserieFin}) by 1. This can be done safely because the contribution of the neglected terms to the asymptotic behavior of $K_N(\alpha)$ decays as least as $(\ln N)^{(\alpha-1)/2}/N^2$, i.e., decays to zero faster than the contribution of the retained terms by (roughly) a factor $N$  [see Eqs.\ (\ref{KNzexp})-(\ref{In2N})].


\begin{table}
\caption{
Parameters appearing in the asymptotic expression of $S_N(t)$,
Eq.\ (\protect{\ref{SNt}}). The symbol $d$D refers to the
$d$-dimensional simple hypercubic lattice. The parameter $\widetilde p$ is
$\left[ 2t(2D\pi)^3/3 \right]^{1/2} p({\bf 0},1)$,
where $p({\bf 0},1)\simeq 1.516386$  \protect{\cite{Hughes}}.
}
\begin{tabular}{cccc}
Case &  $A$ & $\mu$ & $h_1$  \\
\tableline
1D & $\sqrt{2/\pi}$ & 1/2 & -1 \\
2D & $1/\ln t$		&  1  & -1  \\
3D & $1/(\widetilde p\protect{\sqrt{t}})$ &1&-1/3  \\
\end{tabular}
\label{table1}
\end{table}


\begin{figure}
\caption{
The integrand 
$N \, \xi \,\Gamma_{t}^{N-1}\, (d \Gamma_t/d\xi) $
of $J_N(1;0,\infty)$ versus $\xi$ for the one-dimensional lattice and $N=1$, $N=20$ and $N=100$. The solid lines correspond to the integrand when the exact value of $\Gamma_t(\xi)$ is used. The broken lines are obtained by using the first-order asymptotic approximation 
$\Gamma_t(\xi)\approx 1-(2/\pi)^{1/2}\xi^{-1}\exp[-\xi^2/2](1-\xi^{-2})$. 
The filled circle [square] marks the value of $\xi_\times$ for $N=20$ [$N=100$] using  $p=2$ in Eq.\ (\protect{\ref{defxitimes}}).
}
\label{figxitimes}
\end{figure}

\begin{figure}
\caption{A snapshot of the set of sites visited by $N=1000$ random walkers 
on the two-dimensional lattice. 
The visited sites are in white, the unvisited ones
are in black and the internal gray points are the random walkers. The outer 
white circle is centered on the starting point of the random walkers
and its radius is the maximum distance from that point reached by any walker
at the time the snapshot was taken. The internal black circle is
concentric with the former but its radius is the distance between the origin
and the nearest unvisited site.}
\label{figSnapshot}
\end{figure}

\begin{figure}
\caption{
Four successive scaled snapshots of the set of sites visited  by $N=700$ random walkers on the two-dimensional lattice for times (from left to right) $t=2000$, $t=4000$, $t=6000$ and $t=8000$. The second snapshot has been shrunk by the factor 1/$\protect{\sqrt{2}}$, the third  by the factor $1/\protect{\sqrt{3}}$ and the last by the factor $1/2$. 
}
\label{fig:selfsimi}
\end{figure}

\end{document}